\documentclass[12pt]{amsart}

\usepackage{amsmath,amssymb,epic,lscape}

\newtheorem{theorem}{Theorem}[section]

\theoremstyle{definition}

\theoremstyle{remark}

\numberwithin{equation}{section}

\def\DJ{{\hbox{D\kern-.8em\raise.15ex\hbox{--}\kern.35em}}}

\def\SU{{\mbox{\rm SU}}}
\def\Un{{\mbox{\rm U}}}

\def\pH{{\mathcal H}}

\def\pW{{\mathcal W}}

\begin{document}

\title[Dimensions of generic local orbits]
{Dimensions of generic local orbits of multipartite quantum systems}

\author[D.\v Z. \DJ okovi\'c ]
{Dragomir \v Z. \DJ okovi\'c }

\address{Department of Pure Mathematics, University of Waterloo,
Waterloo, Ontario, N2L 3G1, Canada}

\email{djokovic@uwaterloo.ca}

\thanks{The author was supported in part by
the NSERC Grant A-5285.}

\date{}

\begin{abstract}
We consider the action of the group of local unitary transformations,
$\Un(m)\otimes\Un(n)$, on the set of (mixed) states $\pW$
of the bipartite $(m\times n)$ quantum system. We prove that the
generic $\Un(m)\otimes\Un(n)$-orbits in $\pW$ have dimension
$m^2+n^2-2$. This problem was mentioned (and left open) by
Ku\'{s} and \.Zyczkowski in their paper \emph{Geometry of
entangled states} \cite{KZ}.
The proof can be extended to the case of arbitrary finite-dimensional
multipartite quantum systems.
\end{abstract}

\maketitle

\section{Motivation}

We consider the action of the group of local unitary transformations,
$\Un(m)\otimes\Un(n)$, on the set of (mixed) states $\pW$
of the bipartite $(m\times n)$ quantum system.
The Hilbert space of this system has the tensor product structure
$\pH=\pH_A\otimes\pH_B$, with $\pH_A$ of dimension $m$ and 
$\pH_B$ of dimension $n$. We assume that $n\ge m\ge2$.

For the sake of convenience let us denote by ${\rm Herm}(\pH)$ the real vector
space of all hermitian operators on $\pH$. The affine subspace
of ${\rm Herm}(\pH)$ consisting of operators of trace 1 will be denoted
by ${\rm Herm}_1(\pH)$. Note that this affine
space has (real) dimension $m^2n^2-1$.

The set $\pW$ of (mixed) states of our quantum system is the set of
all Hermitian positive semi-definite linear operators on $\pH$
having trace 1. It is a compact convex set with non-empty
interior, relative to the ambient real affine space ${\rm Herm}_1(\pH)$.
Let us recall that the action of the full (global) unitary
group $\Un(mn)$ on the space ${\rm Herm}(\pH)$ is given by
$(U,W)\to UWU^\dag$, where $U\in\Un(mn)$ and $W\in{\rm Herm}(\pH)$.
The action of $\Un(m)\otimes\Un(n)$ is given by
\[ (U\otimes V,W)\to (U\otimes V)\cdot W \cdot (U^\dag\otimes V^\dag), \]
where $U\in\Un(m)$, $V\in\Un(n)$ and $W\in{\rm Herm}(\pH)$.

The problem of computing the maximum dimension of 
$\Un(m)\otimes\Un(n)$-orbits in $\pW$ was raised in a
recent paper of Ku\'{s} and \.Zyczkowski \cite{KZ}.
According to their expectation, this maximum dimension
should be $m^2+n^2-2$. They carried out computations
and verified this claim for some small dimensions.

The object of this note is to prove that the answer to the above
question of Ku\'{s} and \.Zyczkowski is positive for all dimensions.

\section{Result and proof}

We shall prove the following theorem.

\begin{theorem} \label{glavna}
The maximum dimension of local unitary orbits in $\pW$
is $m^2+n^2-2$. The minimal isotropy subgroup for this
action is exactly the center of $\Un(m)\otimes\Un(n)$.
The corresponding orbit is diffeomorphic to the
product manifold $\SU(m)/Z_m\times\SU(n)/Z_n$, where
$Z_m$ resp. $Z_n$ is the center of $\SU(m)$ resp. $\SU(n)$.
\end{theorem}

\begin{proof}
The first and third assertions are consequences of the second. Thus
we shall only prove the second one.

Since the center of $\Un(m)\otimes\Un(n)$ acts trivially on ${\rm Herm}(\pH)$,
it suffices to exhibit a positive semi-definite operator
$W\in{\rm Herm}(\pH)$ whose isotropy subgroup (i.e., the stabilizer) in
$\Un(m)\otimes\Un(n)$ is precisely the center of this group.
In fact we shall drop the condition that $W$ be positive semi-definite
because we can always add to $W$ a positive scalar multiple of
the identity operator to make it positive definite without
affecting its isotropy subgroup.

Let $e_1,\ldots,e_m$ be an orthonormal basis of $\pH_A$
and let $P_1,\ldots,P_m$ be the corresponding orthogonal
projectors, $P_i=e_i e_i^\dag$. Let us introduce an auxiliary vector
$v=\sum ke_k$. We now introduce our Hermitian
operator $W$ by setting
\begin{equation} \label{def-W}
W = \sum_{i=1}^m P_i\otimes X_i + vv^\dag\otimes I,
\end{equation}
where $I$ is the identity operator on $\pH_B$ and
the $X_i$'s are traceless Hermitian operators on $\pH_B$ 
which are chosen so that
their joint centralizer in $\Un(n)$ is just the center of $\Un(n)$,
and the $X_i$'s and $I$ are all linearly independent.
(It is easy to choose such $X_i$'s, and there are
infinitely many such choices.)

Assume that $U\otimes V\in\Un(m)\otimes\Un(n)$ belongs to the
isotropy subgroup of $W$, i.e.,
\[ (U\otimes V)\cdot W\cdot (U^\dag\otimes V^\dag) = W. \]
This can be rewritten as
\begin{equation} \label{jed-W}
W = \sum_i UP_iU^\dag\otimes VX_iV^\dag + Uvv^\dag U^\dag\otimes I.
\end{equation}
Due to the linear independence of the $X_i$'s and $I$ as well as the
linear independence of the $P_i$'s and $vv^\dag$, we can apply a result
from Linear Algebra (see \cite[Chapitre II, \S7, Proposition 17]{NB}
and its first corollary) to deduce from Eqs. (\ref{def-W})
and (\ref{jed-W}) that the operators
$P_i$ and the operator $vv^\dag$ span the same subspace as the operators
$UP_i U^\dag$ and the operator $Uvv^\dag U^\dag$.

Thus, for all $i$'s, we have
\[ UP_i U^\dag=\sum_j \alpha_{ij} P_j + \beta_i vv^\dag, \]
and also
\[ Uvv^\dag U^\dag = \sum_i \gamma_i P_i + \delta vv^\dag, \]
for some real numbers $\alpha_{ij}$, $\beta_i$, $\gamma_i$ and $\delta$.
By plugging these expressions into Eq. (\ref{jed-W}), we obtain
\begin{eqnarray*}
W &=& \sum_i \left( \sum_j \alpha_{ij}P_j + \beta_i vv^\dag \right)
\otimes V X_i V^\dag \\
&& + \left( \sum_j \gamma_j P_j + \delta vv^\dag \right) \otimes I \\
&=& \sum_j P_j \otimes \left( \gamma_j I
+V\left( \sum_i \alpha_{ij}X_i \right) V^\dag \right) \\
&& +vv^\dag\otimes \left( \delta I
+V \left( \sum_i \beta_i X_i \right) V^\dag \right).
\end{eqnarray*}

By comparing this expression with Eq. (\ref{def-W}), we infer that
\[ \delta I+V \left( \sum_i \beta_i X_i \right) V^\dag =I, \]
i.e.,
\[ \sum_i \beta_i X_i +(\delta-1) I = 0. \]
As the $X_i$'s and $I$ are linearly independent, we conclude
that all $\beta_i=0$ and $\delta=1$.

From the above comparison we also obtain the equations
\[ \gamma_j I+V\left( \sum_i \alpha_{ij}X_i \right) V^\dag = X_j. \]
Since all $X_i$'s have trace 0, we deduce that all $\gamma_j=0$.

Thus, for all $i$'s we have 
\[ UP_i U^\dag=\sum_j \alpha_{ij} P_j. \]
For fixed $i$, the $\alpha_{ij}$ are the eigenvalues of $UP_i U^\dag$,
and so exactly one of them is 1 and all other are 0.
Hence, there is a permutation $\sigma$ of $\{1,2,\ldots,m\}$
such that
\[ UP_iU^\dag = P_{\sigma(i)},\quad 1\le i\le m. \]
Consequently,
\begin{equation} \label{druga}
U(e_i)=\xi_i e_{\sigma(i)},\quad |\xi_i|=1,\, 1\le i\le m.
\end{equation}
We also have the equality
\[ Uvv^\dag U^\dag = vv^\dag \]
which implies that $Uv=\xi v$ for some complex number $\xi$
with $|\xi|=1$. By using the definition of $v$, we obtain that
\[ \sum_k k\xi_k e_{\sigma(k)} = \xi \sum_k ke_k. \]
This shows that  $\sigma$ must be the identity permutation
and that all $\xi_i=\xi$.
Hence, $U$ belongs to the center of $\Un(m)$ and $UP_i U^\dag=P_i$
for all $i$'s.

Eqs. (\ref{def-W}) and (\ref{jed-W}) now give
\[ \sum_i P_i\otimes(VX_iV^\dag-X_i) = 0. \]
Since the $P_i$'s are linearly independent,
it follows that $VX_iV^\dag=X_i$ for all $i$.
By our choice of the $X_i$'s,
we conclude that $V$ belongs to the center of $\Un(n)$.

Hence, we have shown that $U\otimes V$ is in the center
of $\Un(m)\otimes\Un(n)$. This concludes the proof of
the theorem.

\end{proof}

\section{Comments}

To illustrate the theorem,
as the basic example one can take $m=n=2$. In that case
$m^2+n^2-2=6$ which is discussed at length in \cite{KZ},
including the orbits of lower dimension.

In the general case,
the orbits having the maximum orbit type (i.e., having the minimum
iso\-tro\-py subgroups, up to conjugacy) are known as orbits of 
\emph{principal type}. It is 
well-known that the union of all orbits of principal type is
an open dense set (see e.g. \cite[Chapter IV, Theorem 3.1]{GB}).
It is also common to refer to the orbits of principal type
as the \emph{generic} orbits. 

We point out that there may exist also non-principal orbits
having the maximal dimension (the same as the principal orbits).
Such orbits are known as \emph{exceptional} (see \cite[p. 181]{GB}).
They are non-trivial covering spaces of the principal orbit.
For instance, in the case of two qubits ($m=n=2$) there exist
at least two types of exceptional orbits. For one of them
the quotient group of the stabilizer modulo the center of
$\Un(2)\otimes\Un(2)$ is $Z_2$ and for the other it is
$Z_2\times Z_2$.

The proof of Theorem \ref{glavna} given above can be extended
to arbitrary multipartite systems.
Let us formulate precisely this generalization.

\begin{theorem} \label{uopstenje}
Let us consider the multipartite quantum system with
$k$ parties with the Hilbert space
\[ \pH=\pH_1\otimes\pH_2\otimes\cdots\otimes\pH_k. \]
Let $d_i=\dim\pH_i$, $i=1,2,\ldots,k$ and assume that
$2\le d_1\le d_2\le\cdots\le d_k$.
Then the maximum dimension of the local unitary orbits
contained in the set $\pW$ of all mixed states of our
quantum system is equal to
\[ d_1^2+d_2^2+\cdots+d_k^2-k. \]
The minimal isotropy subgroup coincides with the center
of the group
\[ \Un(d_1)\otimes\Un(d_2)\otimes\cdots\otimes\Un(d_k) \]
of local unitary transformations and the generic
orbits in $\pW$ are diffeomorphic to the product
manifold
\[ \SU(d_1)/Z_{d_1}\times\SU(d_2)/Z_{d_2}\times
\cdots\times\SU(d_k)/Z_{d_k}, \]
where $Z_{d_i}$, a cyclic group of order $d_i$, is the
center of the special unitary group $\SU(d_i)$.

\end{theorem}

\end{document}